\documentclass{article}

\usepackage[utf8]{inputenc}
\usepackage{lineno} 
\usepackage{xcolor} 
\usepackage{subfigure}
\usepackage{graphicx} 
\usepackage{float} 
\usepackage{booktabs} 
\usepackage{amstext} 
\usepackage[hidelinks]{hyperref} 
\usepackage{pgf} 
\usepackage[numbers]{natbib}

\usepackage{abstract}

\usepackage{amsmath,bm}

\usepackage{enumitem, hyperref}
\newif\ifnewauthor
\newauthortrue

\usepackage{authblk}

\makeatletter

\begin{document}

\begin{sloppypar} 

\title{\textbf{Localise to segment: crop to improve organ at risk segmentation accuracy}}
\date{\vspace{-6ex}} 
\author[1, 2, *]{Abraham George Smith}
\author[1, 2]{Denis Kutnár}
\author[2]{Ivan Richter Vogelius}
\author[1]{Sune Darkner}
\author[1, 2]{Jens Petersen}

\affil[1]{Department of Computer Science, University of Copenhagen}
\affil[2]{Department of Oncology, Rigshospitalet, University of Copenhagen}
\affil[*]{ags@di.ku.dk}

\maketitle

\begin{abstract}
Increased organ at risk segmentation accuracy is required to reduce cost and
complications for patients receiving radiotherapy treatment. Some deep learning
methods for the segmentation of organs at risk use a two stage process where a
localisation network first crops an image to the relevant region and then a
locally specialised network segments the cropped organ of interest. We
investigate the accuracy improvements brought about by such a localisation
stage by comparing to a single-stage baseline network trained on full
resolution images. We find that localisation approaches can improve both
training time and stability and a two stage process involving both a
localisation and organ segmentation network provides a significant increase in
segmentation accuracy for the spleen, pancreas and heart from the Medical
Segmentation Decathlon dataset. We also observe increased benefits of
localisation for smaller organs. Source code that recreates the main results is
available at \href{https://github.com/Abe404/localise_to_segment}{this https
URL}.
\end{abstract}
\section*{Introduction}
\label{sec:intro}

More than 50\% of cancer patients receive radiotherapy which is associated with
a range of dose dependent side effects. Delineation of organs at risk on
treatment planning scans is crucial to minimise
complications \cite{ezzell_guidance_2003, mackie_image_2003}. Manual
delineation is possible and still widely used but in comparison to automated
methods is time consuming \cite{tang_clinically_2019} and  subject to large
inter-observer variation \cite{joskowicz_inter-observer_2019}. Therefore,
methods to improve the accuracy of automated methods are required. A review of
auto-segmentation methods for radiotherapy is presented by
\citet{cardenas_advances_2019} with deep learning methods and convolutional
neural networks in particular representing the state-of-the-art.

Organ localisation has been used for a variety of tasks in image analysis and
can reportedly improve segmentation accuracy whilst reducing computational
memory and processing time requirements \cite{xu_multiple_2019}. 

\citet{kutnar_mixlacune_2021} found a two-stage localisation approach to be
effective for the segmentation of lacunes in brain MR images and
\citet{gros_automatic_2019} found spine centerline localisation to provide
state-of-the-art spinal cord segmentation accuracy. \citet{feng_deep_2019}
proposed a two stage approach using cropped 3D images, where a similar  3D
U-Net was used for both the initial organ localisation and segmentation stages.
They claim their approach is more data efficient due to the use of voxel labels
in the training of the localisation network. The method proposed by
\citet{feng_deep_2019} is appealing as it uses the same method (segmentation
with 3D-U-Net \cite{cicek_3d_2016}) for both localisation and segmentation
which simplifies both concept and implementation.

Although the method obtains competitive accuracy
\cite{yang_autosegmentation_2018}, an ablation analysis or baseline comparison
method is lacking. Therefore, we conduct a more focused investigation to
measure the accuracy gains brought about by such an approach to localisation.
We hypothesise that localisation will improve organ at risk segmentation
accuracy, demonstrated by a significant increase in dice.  To the best of our
knowledge, this hypothesis has not been tested in a focused investigation.

\section*{Method}
\label{sec:methods}

\subsection*{Dataset}
\label{ssec:dataset}

To evaluate the effect of localisation on a diverse array of organ at risk
segmentation tasks, we used the spleen, pancreas, prostate, liver and heart
(left atrium) datasets \cite{simpson_large_2019} from the  Medical Segmentation
 Decathlon \cite{antonelli_medical_2022}.  We used only the original training
sets, as this portion of the data has corresponding labels available for
download.  To facilitate the training and evaluation of deep learning models for image
segmentation, we split the downloaded images and labels into our own training,
validation and test subsets with sizes of 60\%, 20\% and 20\%, respectively
(Table \ref{table:datasize}).  This ratio between training, validation and test
data was chosen as it is typical for deep learning model training. 

   \begin{table}[H]
        \caption{Number of images included in each of the training, validation and test datasets for each of the organs.}
        \label{table:datasize}
        \begin{center}
        \begin{tabular}{@{}llll@{}}
        \toprule
 organ & training & validation & test  \\ \midrule
spleen & 25 & 8 & 8  \\
pancreas & 169 & 56 & 56  \\
prostate & 19 & 7 & 6  \\
liver & 41 & 14 & 14  \\
heart & 12 & 4 & 4  \\
 \bottomrule
            \hline
            \end{tabular}
            \end{center}
            \end{table}

\subsection*{Implementation}
\label{ssec:implementation}

We used PyTorch \cite{paszke_automatic_nodate} (Version 1.13.1) and implemented
a 3D U-Net \cite{cicek_3d_2016} which is an encoder-decoder style
semantic-segmentation architecture. For all experiments we use 64GB of RAM and
two NVIDIA 3090 RTX GPUs.

When performing semantic segmentation using convolutional neural networks, GPU
memory is often a bottleneck. Due to this limitation there is a trade off
between batch size, which is the number of images used in each training update
and patch size, which is the size of the images used during training. Larger
input patches allow more context to be considered for each voxel or pixel
classification decision and have been found to improve accuracy
\cite{huang_tiling_2019}. Therefore we used an input patch size of 64x256x256
for all experiments as this was the largest we could fit in GPU memory. However
as such large input patches take up more GPU memory they force a reduction in
batch size. Therefore we used a batch size of 2 for all experiments with one
instance (input patch) on each GPU, utilising a data-parallel approach, meaning
the training batch is split accross the GPUs.  Small batch sizes can be
problematic for the commonly used batch normalisation method
\cite{ioffe_batch_2015, wu_group_2018}. Therefore we used group normalisation
\cite{wu_group_2018} after each layer as it performs well when small batch
sizes are used and has been found to be effective for 3D medical image
segmentation tasks \cite{myronenko_3d_2019}.

We use a loss function which is a combination of dice
\cite{sudre_generalised_2017} and cross-entropy as this has been found to be
effective when dealing with class imbalanced datasets
\cite{smith_segmentation_2020, taghanaki_combo_2018}.  Although in
\cite{feng_deep_2019} the authors used cross-entropy with importance weights
for their main experiments, they mentioned that they also found a combination
of cross-entropy and dice loss to both stabilise and accelerate the training
process. Another disadvantage of cross-entropy as opposed to dice loss is that
organ specific importance weights require manual tuning.
We used zero padding in the convolution operations to allow our 3D network to
produce an output segmentation with the same size as the input patch.

For all experiments we used the Adam optimiser \cite{kingma_adam_2017} with a
learning rate of 0.0001. For each training run we initialise the
weights using He \cite{he_delving_2015} initialisation. We used check-pointing
and early stopping \cite{morgan_generalization_1990} to mitigate over-fitting.
Check-pointing involves saving the model weights to disk during the training
run. We computed the dice on the validation set at the end of each epoch
and only saved models which obtained a new highest dice. There are
various way to implement an early stopping procedure
\cite{prechelt_early_nodate}. Our stopping criterion used the number of epochs
since an improved dice had been found, a parameter which is a commonly
referred to as patience. We set patience to 20 for all experiments, thus each
training run would stop after 20 epochs had passed since a new highest dice
score on the validation set had been obtained.

To mitigate the possibility that the results were due to chance,  for each
organ and method, training runs were repeated until 10 runs had converged,
where convergence was defined as the model having at least 0.1 dice on the
validation set after 20 epochs. 

\begin{figure}[H]
\thispagestyle{empty}

 \hspace{-0.25cm}
  \makebox[\textwidth][c]{\scalebox{0.40}{\includegraphics{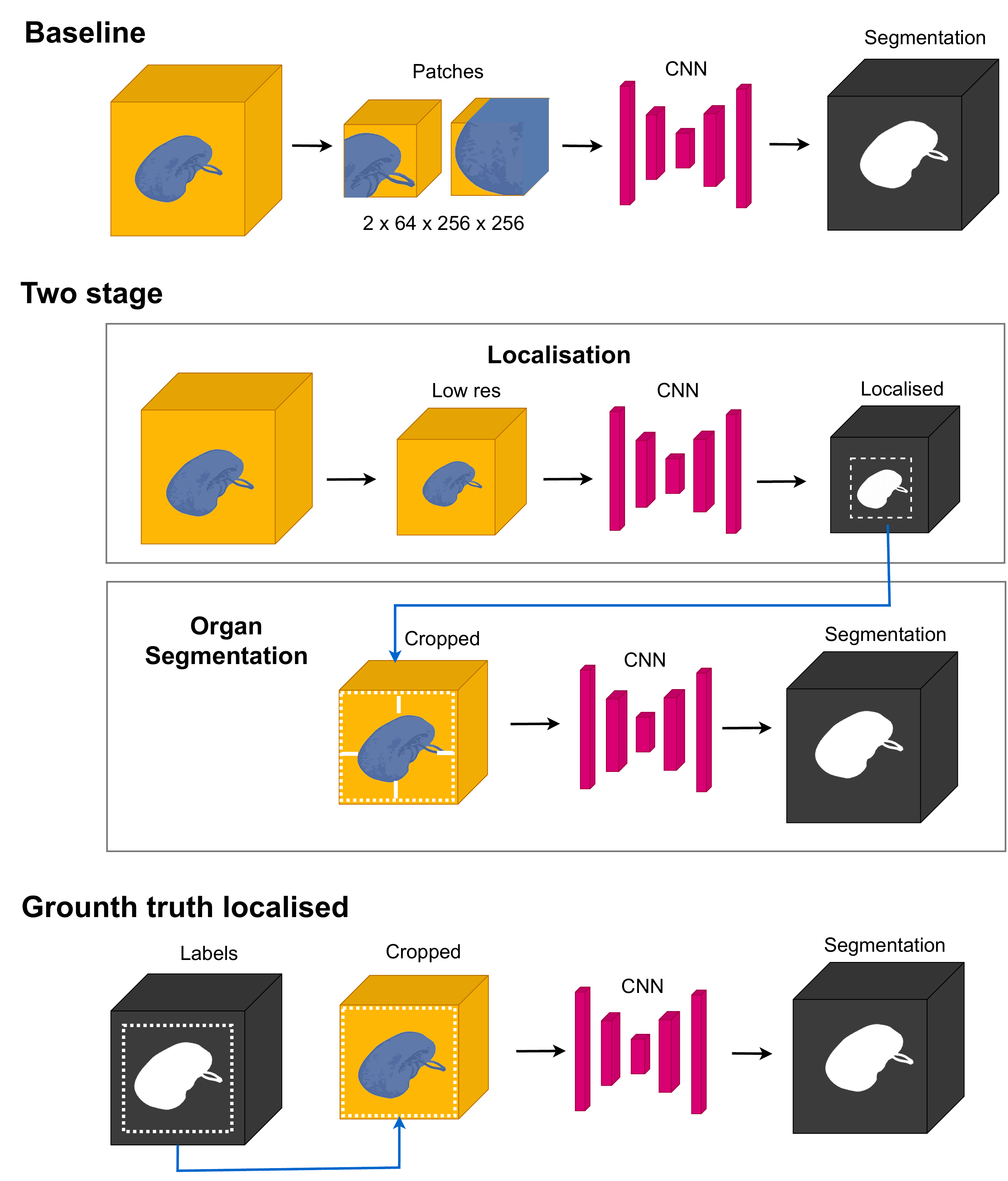}}}
  \caption{Illustration showing the three different methods compared, including the baseline, 
    two stage involving both a localisation network and organ segmentation network and the organ segmentation network that uses the ground truth to localise.}\label{fig:methods}
\end{figure}

The methods campared include a baseline full resolution segentations approach
using 3D patches, a two stage localisation approach and a method involving only
organ segmentation, where the ground truth was used to localise
(Figure~\ref{fig:methods}). In the following sections we describe the three
different approaches we experimented with to evaluate the benefits of
localisation.

\subsection*{Baseline - full resolution segmentation}

In order to evaluate the advantages of the two stage localisation process we
trained a single stage baseline network. For each training instance we sample a
patch with random location within the image and the corresponding location from
the annotation.
We enforced that at least 80\% of the selected patches contained foreground
annotation. Such biased instance selection is a relatively common practice as
otherwise most patches would not contain foreground which can cause convergence
problems.

\subsection*{Localisation network}

In order to train the localisation network we created a low resolution version
of the dataset by resizing the images and annotations down to a half their
width and height and a third of their depth. We then trained the network to
predict the annotations which were also resized to match the reduced resolution
images.  We created these low res images using the resize function from
scikit-image \cite{walt_scikit-image_2014} (Version 0.17.2).

\subsection*{Organ segmentation network}

To train the organ segmentation network, we first created a dataset of images
and annotations which were cropped by taking the region of the image including
the organ with 15 voxels padding on each side to include some background
context. To ensure enough padding was included on each side of the organ, even
if the organ was at the edge of an image, the images were zero padded by 15
voxels on each side before cropping to the organ. The organ segmentation
network was trained independently using these cropped versions of the original
images and ground truth annotations, without regard to the output of any
particular localisation network.

\subsection*{Ground truth localisation}

We also evaluated an approach using a localisation stage which utilises the
ground truth labels. We do this to access the advantages of localisation given
an accurate bounding box.

\subsection*{Two stage localisation \& segmentation pipeline}

In order to segment the full resolution image with a preliminary localisation
step we implemented a two stage process. We first computed a low res version of
the image and then segmented it using the localisation network. We then
identified the organ as the largest connected foreground region in the low res
segmentation. We then segmented the corresponding region in the full resolution
image with padding on each side of the organ as described in the above \textit{Organ
segmentation network} section. To perform this two stage segmentation, we pairwise
couple the localisation networks with the organ segmentation networks
chronologically, thus the i'th organ network trained is coupled with the i'th
localisation network.

\subsection*{Metrics}
During training we computed dice on both the training and validation
data. For the final model that was automatically selected at the end of the
training run, dice was computed on both the randomly selected validation and
test sets using the full resolution segmentations and annotations.

The two sided t-test as implemented in SciPy \cite{virtanen_scipy_2020}
(Version 1.5.4) was used for testing for significant differences between the
accuracy of the methods on both the validation and test datasets. We also record time for 
each of the training runs to converge.

\section*{Results}
\label{sec:results}

\subsection*{Training}

\begin{figure}
\vspace{-4cm}
  \hspace{-0.38cm}
  \makebox[\textwidth][c]{\scalebox{0.257}{\input{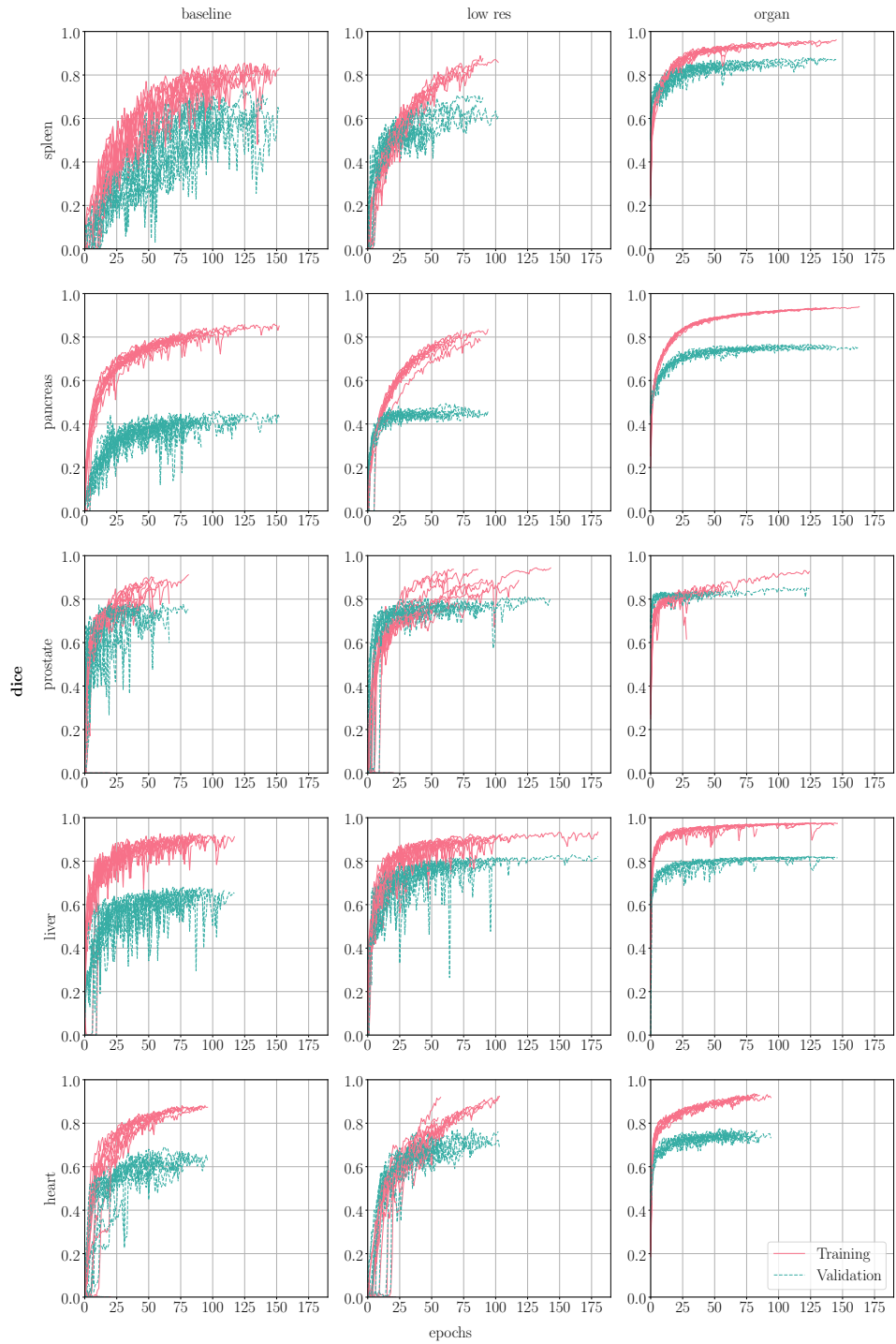}}}
  \caption{Validation and training dice are shown for each epoch for each of the 10 training runs for each organ and for each method, resulting in 150 training runs in total. Only training runs which successfully converged are shown.}
  \label{fig:training_stability}
\end{figure}

\begin{figure}
  \makebox[\textwidth][c]{\scalebox{0.445}{\input{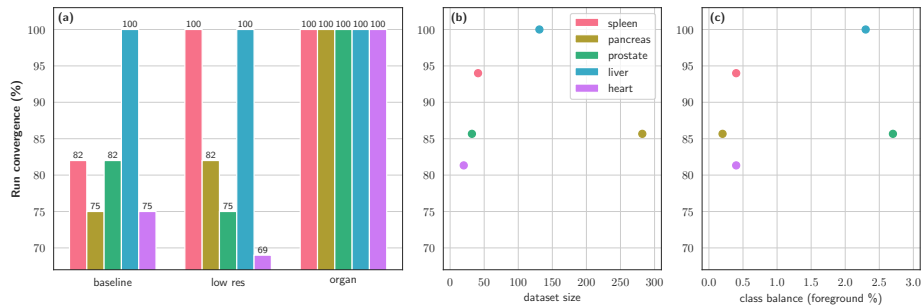}}}
  \caption{ Convergence rate for (a) each method and organ, (b) as a function of dataset size and (c) as a function of class balance (average foreground percent).}
  \label{fig:convergence_stats}
\end{figure}

\begin{table}[H]
\caption{
\label{table:traintime}
Average training time (minutes) for each of the three types of networks for each of the organs.
}
\begin{center}
\begin{tabular}{@{}llll@{}}
\toprule
 & baseline & low res & organ \\ \midrule
spleen & 74.9 & 10.8 & 10.1 \\
pancreas & 477.6 & 65.1 & 91.2 \\
prostate & 6.4 & 2.0 & 2.2 \\
liver & 253.2 & 42.8 & 110.0 \\
heart & 22.2 & 3.7 & 2.5 \\ \bottomrule
\hline
\end{tabular}
\end{center}
\end{table}

For each organ, the baseline approach is substantially slower than the other
methods and for all organs takes longer to converge than both the low res and
organ networks combined (Table \ref{table:traintime}).

Both the baseline method and low res networks had less stable performance
during training compared to the organ network, with larger fluctuations in the
dice (Figure \ref{fig:training_stability}). The organ segmentation network
always converged (Figure \ref{fig:convergence_stats}a).  With the baseline and
low res networks, convergence is similarly likely, with 83\% of the baseline
training runs and 85\% of the low res training runs converging. The varying
rates of convergence (Figure \ref{fig:convergence_stats}) reflect the
difference seen in training stability (Figure \ref{fig:training_stability}).

\subsection*{Validation}

\begin{table}[H]
\caption{\label{table:valdice}
Average dice on the validation set for the baseline network compared to the two stage approach with both predicted and ground truth localisation. Values which are significantly higher than the baseline are shown in bold.}
\begin{center}
\begin{tabular}{@{}llll@{}}
\toprule
           & baseline & two stage & ground truth localised \\ \midrule
    spleen & 0.6491 $\pm$ 0.0997 & $\pmb{0.8142 \pm 0.0221}$ & $\pmb{0.8619 \pm 0.0092}$ \\
    pancreas & 0.4372 $\pm$ 0.0148 & $\pmb{0.6674 \pm 0.0146}$ & $\pmb{0.7564 \pm 0.0096}$ \\
    prostate & 0.7744 $\pm$ 0.0109 & 0.7699 $\pm$ 0.0149 & $\pmb{0.8323 \pm 0.0076}$ \\
    liver & 0.6661 $\pm$ 0.0188 & 0.7044 $\pm$ 0.0694 & $\pmb{0.7807 \pm 0.1061}$ \\
    heart & 0.6547 $\pm$ 0.0239 & $\pmb{0.7018 \pm 0.0281}$ & $\pmb{0.7612 \pm 0.0087}$ \\
\bottomrule
\hline
\end{tabular}
\end{center}
\end{table}

For the validation sets, the dice was significantly higher for the two stage
approach compared to the baseline method for the heart ($p < 0.001$), spleen
($p < 0.001$) and pancreas ($p < 0.001$). For the liver, although the two stage
approach appears it may offer some improvements, the difference was not
significant ($p=0.11$). For the prostate there was no significant difference
($p=0.44$).

%

On the validation set, the difference between the ground truth localised method
and baseline was significant for the liver ($p < 0.05$) and highly significant
for the heart, spleen, pancreas and prostate ($p < 0.001$). For all organs
except the liver ($p=0.07$), the benefits of ground truth localization are
significant compared to using the localization network to provide the cropped
region ($p < 0.001$).

%

\subsection*{Test}

\begin{table}[H]
\caption{\label{table:testdice}
Average dice on the test set for the baseline network compared to the two stage approach with both predicted and ground truth localisation. Values which are significantly higher than the baseline are shown in bold.}
\begin{center}
\begin{tabular}{@{}llll@{}}
\toprule
         & baseline & two stage & ground truth localised \\ \midrule
spleen   & 0.4433 $\pm$ 0.1162 & $\pmb{0.6503 \pm 0.0538}$ & $\pmb{0.8255 \pm 0.0086}$  \\
pancreas & 0.4366 $\pm$ 0.0205 & $\pmb{0.6519 \pm 0.0136}$ & $\pmb{0.7397 \pm 0.0063}$  \\
prostate & 0.797 $\pm$ 0.0342 & 0.7204 $\pm$ 0.0244 & $\pmb{0.8361 \pm 0.0145}$  \\
liver    & 0.6039 $\pm$ 0.0214 & 0.6096 $\pm$ 0.0664 & $\pmb{0.7156 \pm  0.1028}$  \\
heart    & 0.5764 $\pm$ 0.0311 & $\pmb{0.623 \pm 0.0523}$ & $\pmb{0.7508 \pm 0.0178}$  \\ \bottomrule
\hline
\end{tabular}
\end{center}
\end{table}

On the test set, the two stage dice score was higher than the baseline for the
spleen ($p < 0.001$), pancreas ($p < 0.001$) and heart ($p < 0.05$). For the
liver the difference was not significant ($p = 0.8$). The test set dice 
was significantly higher with the baseline approach compared to the two stage
method for the prostate ($p < 0.001$).


When using the ground truth labels to localise, the increase in organ network
dice compared to the baseline was highly significant for the heart, spleen and
pancreas ($p < 0.001$) and significant for the prostate and liver ($p < 0.05$).

\begin{figure}[H]
    \begin{center}
        \scalebox{0.7}{\input{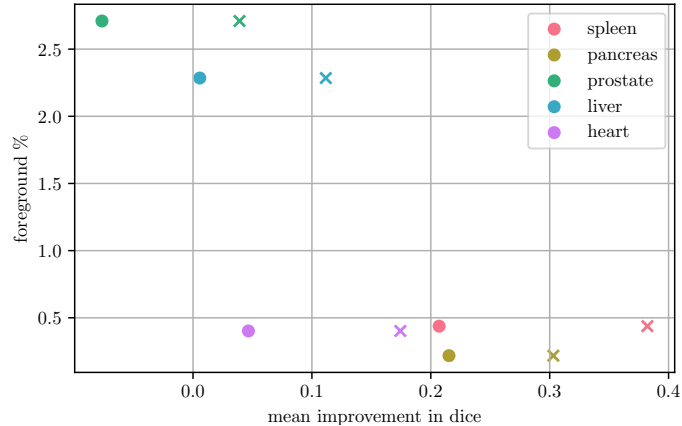}}
\caption{Benefit of localisation for each of the datasets for both predicted
localisation region (o) and when the ground truth location is provided (x). The
mean improvement in dice is calculated by subtracting the mean baseline dice from
the mean localised dice. Foreground \% is the percentage of the voxels in a
scan that belong to the organ as opposed to the background, where background
is considered as all voxels outside of that particular organ.}
\label{fig:loc_benefit} \end{center} \end{figure}

The benefits of ground truth localization were also highly significant compared
to using the localization network to provide the cropped region for the heart,
spleen, pancreas and prostate ($p < 0.001$) and significant for the liver ($p <
0.05$). We found that smaller organs, as a percentage of scanned region tend to
benefit more from localisation (Figure \ref{fig:loc_benefit}).

%

\section*{Discussion \& Conclusion}
\label{sec:discussion}

Although the significant improvements in dice for the majority of
datasets confirm our hypothesis that localisation improves organ at risk
segmentation accuracy, the baseline performed stronger than expected in
comparison to the two stage localisation approach, even out-performing the
localisation approach on the prostate test set. 

The mean organ volume as a percentage of total image volume ranges from 0.2\%
for the pancreas to 2.7\% for the prostate. This represents an extreme class
imbalance, particularly for the pancreas, spleen and left atrium. Class
imbalance is known to have detrimental effects on the performance of machine
learning models \cite{ling_class_2010} and convolutional neural networks in
particular \cite{buda_systematic_2018}. If not addressed, a class imbalance
problem may lead to algorithms tending to predict only the majority class
\cite{drummond_severe_2005}.  \citet{gros_automatic_2019} argue a two stage
approach involving localisation is able to mitigate issues caused by class
imbalanced data. The trend of an increased benefit of localisation for smaller
organs (Figure \ref{fig:loc_benefit}) is expected, because for smaller organs
the class balance issue becomes more severe and if the organ becomes large
enough there will be negligible difference between the baseline and
localisation approaches.

One explanation for the good performance of the baseline could be the random
selection of patches during the baseline training procedure. This random
selection could have provided some augmentation benefits similar to
random cropping. When the organ segmentation network encountered unexpected
anomalies it may have been less equipped to handle them.
\citet{xu_efficient_2019} trained an organ segmentation network using a region
containing the organ of interest but with variation in the amount of padding
around it. Varying the amount of context around each organ during training may
be key to a two-stage localisation network that provides consistent advantages
in accuracy compared to the single stage baseline line method.

The baseline and low res network training instability, including fluctuations
in dice (Figure \ref{fig:training_stability}) is likely related to the
challenges with class imbalance. Although the baseline network had biassed
instance selection to include foreground batches more frequently, its task was
likely more complicated compared to cropped organ segmentation as the baseline
network must learn to segment regions further away from the organ. For the
baseline approach, the patches used in training will have also been less
consistent, including varying amounts of the organ of interest or sometimes
only background regions.

The heart (left atrium) had the lowest rate of convergence on average, which is
likely due to it having both a relatively small dataset (Figure
\ref{fig:convergence_stats}b) and a large class imbalance (Figure
\ref{fig:convergence_stats}c). Our condition for convergence was based on
accuracy, which typically increases with training dataset size
\cite{henderson_impact_2023}. An exception is the pancreas, with the largest
dataset, yet a convergence rate of only around 85\% (Figure
\ref{fig:convergence_stats}b), which may be due to the high class imbalance in
this dataset (Figure \ref{fig:convergence_stats}c).

Reduction in model training time is critical for both workflow optimisation and
carbon footprint \cite{anthony_carbontracker_2020}.  Slow training may also
hinder novel interactive-machine-learning approaches that depend on model
adaptation to support a feedback loop between annotator and model
\cite{smith_rootpainter3d_2022, smith_rootpainter_2022, wei_towards_2023,
smith_pd-0065_2022}. We found that the baseline method had slower convergence
and longer training time compared to the localisation approach, even when
considering that the localisation approach involved training two 
networks (Table \ref{table:traintime}). The slow convergence of patch based
training in comparison to other approaches has been observed in previous work
evaluating methods for brain tumour segmentation \cite{bouget_meningioma_2021}.

A potential drawback of the two stage localisation approach is the additional
complexity of training two networks. A potential limitation to the network
architecture used in this study is the use of zero-padding to ensure that the
network input and output had consistent size.  In some cases zero-padding has
been found to increases errors on the edge of a patch by as much as 35\%
\cite{huang_tiling_2019}. 

The consistent benefits of using ground truth to localise the region for the
organ segmentation network (Table \ref{table:testdice}) motivate the use of a
manual bounding box in cases where accuracy improvements are required for
smaller organs, an approach that has been used for prior studies in interactive
machine learning for organ at risk segmentation
\cite{smith_rootpainter3d_2022}. Manual localisation would also be feasible
with interactive segmentation methods, such as the approach proposed by
\citet{rasmussen_simple_2023} where organ extremities are input to
guide the predicted contour.

Our results show the advantages of both manual and automatic localisation
for organ at risk segmentation in terms of both training time, convergence rate
and segmentation accuracy, especially for smaller organs where class imbalance
causes challenges for conventional approaches to segmentation model training.

\bibliography{references}
\bibliographystyle{plainnat}
\end{sloppypar}
\end{document}